\documentclass[12pt]{article}

\usepackage{epsfig}
\usepackage{setspace}
\usepackage{subfigure}

\usepackage{graphicx}
\usepackage{longtable}
\usepackage{amsmath}

\def\e{\begin{equation}}
\def\f{\end{equation}}
\def\%#1{\mbox{\boldmath $#1$}}
\def\=#1{\overline{\overline #1}}

\def\_#1{{\bf #1}}

\def\E{\epsilon}
\def\M{\mu}

\def\.{\cdot}

\def\##1{{\bf#1\mit}}

\def\l#1{\label{eq:#1}}
\def\r#1{(\ref{eq:#1})}
\def\am{\left(\begin{array}{c}}
\def\amm{\left(\begin{array}{cc}}
\def\a{\end{array}\right)}

\title{On Artificial Magneto-Dielectric Loading for Improving the Impedance Bandwidth Properties of Microstrip Antennas}

\author{Pekka~Ikonen, Stanislav Maslovski, Constantin Simovski, and Sergei Tretyakov}

\date{Radio Laboratory/SMARAD, Helsinki University of
Technology\\P.O. Box 3000, FI-02015 TKK, Finland}

\begin{document}

\maketitle {\center \large

Address for correspondence:

Pekka Ikonen, \\ Radio Laboratory, Helsinki University of
Technology,\\ P.O. Box 3000, FI-02015 TKK, Finland.

Fax: +358-9-451-2152

E-mail: pikonen@cc.hut.fi

}

\parindent 0pt
\parskip 7pt

\vspace{0.5cm}

\begin{center}
\section*{Abstract}
\end{center}

In the present paper we discuss the effect of artificial
magneto-dielectric substrates on the impedance bandwidth properties
of microstrip antennas. The results found in the li\-te\-ra\-tu\-re
for antenna miniaturization using magnetic or magneto-dielectric
substrates are revised, and discussion is addressed to the
practically realizable artificial magnetic media operating in the
microwave regime. Using a transmission-line model we, first,
reproduce the known results for antenna miniaturization with
non-dispersive material fillings. Next, a realistic dispersive
behavior of a practically realizable artificial substrate is
embedded into the model, and we show that frequency dispersion of
the substrate plays a very important role in the impedance bandwidth
characteristics of the loaded antenna. The impedance bandwidths of
reduced size patch antennas loaded with dispersive
magneto-dielectric substrates and high-permittivity substrates are
compared. It is shown that unlike substrates with dispersion-free
permeability, practically realizable artificial substrates with
dispersive magnetic permeability are not advantageous in antenna
miniaturization. This conclusion is experimentally validated.

\textbf{Key words}: Patch antenna, high-permeability materials,
magneto-dielectric substrate, frequency dispersion, quality factor,
miniaturization.

\section{Introduction}

For a long time microstrip antennas have been miniaturized using
different material fillings between the radiating element and the
ground plane \cite{Bahl,Balanis,Pozar}. Most traditionally, high
permittivity dielectrics have been used to decrease the physical
dimensions of the radiator \cite{Hwang,Colburn}. Common problems
encountered with high permittivity substrates include e.g.~the
excitation of surface waves leading to lowered radiation efficiency
and pattern degradation, and difficulties in impedance matching of
the antenna. During the first heyday of so called metamaterials,
artificial high-permeability materials working in the microwave
regime have gained increasing attention
\cite{Maslovski}--\cite{Simovski_surface}. The possibility to create
artificial magnetism at microwave frequencies has heated the
discussion on the possibility to enhance the impedance bandwidth
properties of planar radiators using magneto-dielectric substrates
\cite{Yoon}--\cite{Karkkainen_MOTL} or other electromagnetically
exotic substrates \cite{Sievenpiper,Mosallaei,Zhao}.

According to the work of Hansen and Burke \cite{Hansen}, inductive
(magnetic) loading leads to an efficient size miniaturization of a
microstrip antenna. A transmission-line (TL) model for a normal
half-wavelength patch antenna predicts that increase in the
permeability of the antenna substrate does not reduce the impedance
bandwidth of the miniaturized radiator (when the material parameters
are \emph{dispersion-free}, and $\mu_{\rm eff}\gg\E_{\rm eff},
\mu_{\rm eff}\gg1$) \cite{Hansen}. Edvardsson \cite{Edvardsson}
derived a condition for the radiation quality factor $Q_{\rm r}$ of
a planar inverted F-antenna (PIFA), and concluded that increase in
the permeability allows size miniaturization without increasing
$Q_{\rm r}$. However, to draw a general conclusion on the practical
applicability of the magnetic loading e.g.~with mobile phone
antennas, there are three things which need to be clarified: i) The
effect of frequency dispersion of the substrate. Practically
realizable artificial magnetic medium operating at microwave
frequencies is composed of electrically small resonating metal unit
cells\footnote{Paramagnetic response at microwave frequencies can
also be achieved using hexaferrites, or composite materials
containing ferromagnetic inclusions. In the present paper we
consider, however, only artificial magnetic media composed of
resonating metal inclusions.}, thus, the magneto-dielectric
substrate obeys strong frequency dispersion. Even though the
substrate can be considered as a paramagnetic medium when using
plane wave excitation (at the macroscopic level, over a certain
frequency range), the performance is not so obvious to predict when
the resonant substrate is coupled to the antenna element. ii) The
effect of high $\E_{\rm eff}$ compared to $\mu_{\rm eff}$ over the
matching band of the antenna. With practically realizable artificial
magneto-dielectric substrates for microwave frequencies
Re$\{\mu_{\rm eff}\}$ is known to be rather moderate, thus in
practice it seems very difficult to achieve the condition $\mu_{\rm
eff}\gg\E_{\rm eff}, \mu_{\rm eff}\gg1$ (outside the resonant region
of the substrate). iii) The effect of losses. Realistic model
representing the losses of the substrate is needed to correctly
predict the behavior of the susceptance seen at the antenna terminal
(further, the quality factor). Moreover, the numerical and
experimental results found in the literature
\cite{Yoon}--\cite{Karkkainen_MOTL} do not offer a complete,
quantitative comparison scheme applicable for practical antenna
design. According to the authors knowledge, the impedance bandwidths
(quality factors) of reduced size antennas loaded with practically
realizable, artificial, dispersive magneto-dielectrics have not been
compared to the results obtained using high-permittivity dielectrics
leading to the same size reduction. When thinking of practical
antenna design, this quantitative comparison is clearly needed to
make decisions whether or not the possibly enhanced impedance
bandwidth outweights e.g.~the increased manufacturing cost and
substrate weight.

In real life large permeabilities for microwave materials are
achieved with a complex mixture of electrically small
inhomogeneities (resonating unit cells) loaded into a substrate in a
specific periodic arrangement. In the present paper we shortly
discuss the possibilities to enhance artificial magnetism at
microwave frequencies, and construct a TL model taking into account
realistic dispersive behavior of a practically realizable artificial
magneto-dielectric substrate. We start the analysis by reproducing
the known miniaturization results with static material parameters,
and further extend the study to take into account frequency
dispersion. It is shown that a substrate obeying the Lorentzian type
dispersion for magnetic permeability leads to a narrower impedance
bandwidth than a high-permittivity substrate offering the same size
reduction. Physical reason for the phenomena is clarified. A
prototype antenna is constructed and the observation is
experimentally validated.

The rest of the paper is organized in the following way: In section
II we briefly revise the numerical and experimental results found in
the literature for magneto-dielectric loading of microstrip
antennas. Section III presents a short discussion on practically
realizable artificial magnetic media and the used TL-model is
presented in detail in Section IV. Section V presents the calculated
impedance bandwidth properties in different loading scenarios, and
an experimental demonstration is presented in section VI. The work
is concluded in Section VII.

\section{Revision of numerical and experimental results}

Mosallaei and Sarabandi applied in \cite{Mosallaei_disagree} a
finite-difference time-domain (FDTD) simulation scheme and
investigated antenna miniaturization using a band-gap substrate
consisting of dielectric and magneto-dielectric (Z-type hexaferrite)
layers. Constant scalar-permeability assumption was used with
substrate material parameters $\E_{\rm eff}=3.84(1 - j0.001)$,
$\mu_{\rm eff}=8.61(1 - j0.018)$ at 277 MHz. Magneto-dielectric
substrate was shown to lead to a significantly wider impedance
bandwidth than a pure dielectric substrate offering the same size
reduction. The same authors introduced in \cite{Mosallaei_Pisa} an
embedded circuit metamaterial structure and utilized it under a
patch antenna operating at 2.3 GHz (FDTD simulations). A 1.5 percent
fractional bandwidth was demonstrated for a 0.075$\lambda_0$ size
patch antenna with lossless, dispersion-free substrate material
parameters ($\mu_{\rm eff} \simeq 3.1, \E_{\rm eff} \simeq 9.6$).
The authors did not, however, present any result for high
permittivity dielectrics leading to the same size reduction. Buell
\emph{et.~al} \cite{Buell2} measured the performance of a patch
antenna utilizing an embedded circuit metamaterial at 250 MHz, and
reported that miniaturization factors as high as 6.4 (with $BW|_{\rm
-10 dB}$ 0.83 percent) could be achieved. However, the radiation
efficiency was measured to be only 21.6 percents, and a comparative
measurement with high-permittivity dielectrics was not conducted.

Yoon and Ziolkowski \cite{Yoon} simulated a microstrip antenna with
different material fillings (dispersion-free material parameters),
and came to the conclusion that the optimal condition for the
effective substrate material parameters is the same as predicted by
Hansen and Burke \cite{Hansen},  and Edvardsson \cite{Edvardsson}
($\E_{\rm eff}=1, \mu_{\rm eff}\gg1$). In \cite{Karkkainen}
K\"arkk\"ainen \emph{et.~al} numerically studied a PIFA with
dispersive magnetic material filling. The authors pointed out the
need to regard the loaded radiator as a system of two coupled
resonators, rather than a radiator loaded with a static paramagnetic
load. The conclusion was that if the resonance of the material is
considerably higher than the resonance of the empty antenna, the
material loading enables size reduction while approximately
retaining the fractional bandwidth. The effective permittivity of
the substrate was assumed to be unity, and the design utilizing
dispersive magnetic filling was not challenged against a design
utilizing dispersion-free dielectrics offering the same size
reduction.

Ermutlu \emph{et.~al} \cite{Ermutlu} loaded the volume under a
half-wavelength patch antenna with a practically realizable
artificial magneto-dielectric material, and concluded based on
numerical and experimental results that significant miniaturization
factors could be achieved while the bandwidth of the loaded antenna
is approximately retained. The results were not, however, challenged
against results obtained using high-permittivity substrates. Authors
of \cite{Pekka_MOTL} proposed two utilization schemes for a resonant
magnetic media with planar radiators. The volume under a PIFA was
loaded with an array of metasolenoids \cite{Maslovski} and the
impedance bandwidth properties were experimentally investigated.
When utilizing the resonant region of the metasolenoids, the authors
designed a multiresonant antenna with significantly enhanced
impedance bandwidth. According to the second suggestion, the
metasolenoid array would behave as a paramagnetic load enabling
efficient antenna miniaturization\footnote{The figure of merit used
to evaluate the performance of a PIFA with different material
fillings is incorrectly calculated in \cite{Pekka_MOTL}. When the
figure of merit is calculated using radiation quality factor $Q_{\rm
r}$, air filling leads to the best figure of merit.}. Authors of
\cite{Karkkainen_MOTL} loaded the volume under a half-wavelength
patch antenna with an array of metasolenoids and compared the
impedance bandwidth to that obtained using high-permittivity
dielectrics. The result indicated that there is practically no
advantage when using the metasolenoid array.

Attempts have also been conducted to miniaturize the size of planar
antennas by partially filling the volume under the radiating element
with backward-wave materials \cite{Mahmoud,Tretyakov}. With patch
antennas, in practise, filling the volume with backward-wave
material corresponds to inserting bulk inductors and capacitors to
the antenna \cite{Tretyakov}, thus the technique is very similar to
the well known technique of size miniaturization using reactive
loads. Moreover, as expected, material obeying realistic dispersive
behavior results in significantly narrower impedance bandwidth than
the hypothetical material with $\E_{\rm eff}=\mu_{\rm eff}=-1$
\cite{Tretyakov}.

\section{On practically realizable artificial magnetic media at microwave frequencies}

In the present paper we restrict the discussion only to artificial
magnetic media composed of resonating metal unit cells. Other ways
to enhance the magnetic response include e.g.~the utilization of
ferrites, or composites filled with ferromagnetic inclusions
\cite{Lagarkov1, Lagarkov2}. In general, resonant artificial
magnetic media can be utilized  with planar radiators in two
principal ways \cite{Pekka_MOTL}: If the resonance of the material
lies inside the operational band of the (loaded) radiator, with a
suitable coupling the resonance of the material can be combined with
the antenna resonance, thus a multiresonant antenna is achieved. In
other words, radiating particles are used instead of a real material
filling. Another way is to design the material to resonate at a
considerably higher frequency than the operational frequency of the
loaded radiator. In the latter case it is important that the
utilized material retains its effective magnetic properties over a
wide frequency band, meaning that inside the operational band of the
radiator the value of the real part of the effective permeability is
larger than unity, Re$\{\mu_{\rm eff}\}>1$ (low losses are another
desired feature). This allows one to consider the material as a
paramagnetic load with (almost) constant real part for the effective
permeability over the desired frequency range. A paramagnetic load
will increase the inductance of the resonator, and many authors
consider this as an advantage when trying to retain the impedance
bandwidth characteristics in antenna miniaturization with material
fillings, see e.g.~\cite{Yoon, Hansen, Edvardsson}.

The problem with most of the introduced designs for artificial
magnetic media operating in the microwave regime
\cite{Maslovski}--\cite{Simovski_surface}
 is the fact that
the effective magnetism rapidly vanishes as the frequency deviates
from the particle resonance. Usually the maximum predicted value for
the real part of the effective permeability of artificial magnetic
medium (with realistic bulk concentration/volume filling ratio and
loss factor) is Re$\{\mu_{\rm eff}\}=1.5-4$ at the resonance
\cite{Kostin, Pendry, Lagarkov, Buell2, Sim}. This means that in
practice Re$\{\mu_{\rm eff}\}$ of the magneto-dielectric substrate
approaches unity very quickly as the frequency deviates from the
material resonance. So in practice one should utilize the particles
rather close to their resonance to achieve a condition Re$\{\mu_{\rm
eff}\}>1$. This further increases the effect of frequency dispersion
in the impedance bandwidth characteristics. Moreover, it might not
be clear anymore if the loaded antenna should be considered as a
system of two coupled resonators rather than a resonator filled with
a homogenous, dispersion-free material. In addition to this, usually
the real part of the effective permittivity of magneto-dielectric
substrates is considerably higher than the real part of the
effective permeability (when the substrate is utilized well below
its resonance). Thus, at microwave frequencies the condition
considered e.g.~by Hansen and Burke ($\mu_{\rm eff}\gg\E_{\rm eff},
\mu_{\rm eff}\gg1$) \cite{Hansen} seems unlikely to be achieved with
a resonant media composed of metal unit cells. For example, with the
structure presented in \cite{Simovski_surface} a realistic value for
the real part of the effective permeability over the operational
band of a patch antenna (operating well enough below the
magneto-dielectric substrate resonance) is Re$\{\mu_{\rm
eff}\}\simeq1.5$ \cite{Sim}. However, at the same time the real part
of the effective permittivity can be as high as Re$\{\E_{\rm
eff}\}\simeq8.0$, even if the host substrate has a low value for the
relative permittivity \cite{Sim}.

\begin{figure}[b!]
\centering \epsfig{file=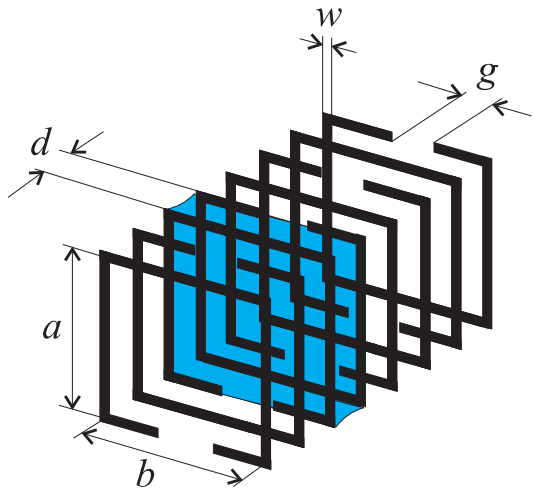, width=7cm} \caption{A
schematic illustration of the metasolenoid.} \label{meta_meas}
\end{figure}

To give a link between the presented effective medium model and
practical design for artificial magnetic media, we need to choose a
unit cell for the media. In the present work we consider the
metasolenoid \cite{Maslovski,Liisi} as an example unit cell for
practically realizable artificial magnetic media.
Fig.~\ref{meta_meas}.~introduces the structural geometry of the
metasolenoid. Metasolenoid is chosen due to its simple structural
geometry enabling rather straightforward, yet accurate theoretical
analysis. Moreover, experimental results can be found in the
literature for the utilization of the metasolenoid under microstrip
antennas \cite{Pekka_MOTL, Karkkainen_MOTL}, and the effective
medium model has been experimentally validated and found to be
accurate \cite{Maslovski}.

\section{Transmission-line model}

\label{tl_model_formulation} 
\begin{figure}[b!]
\centering \epsfig{file=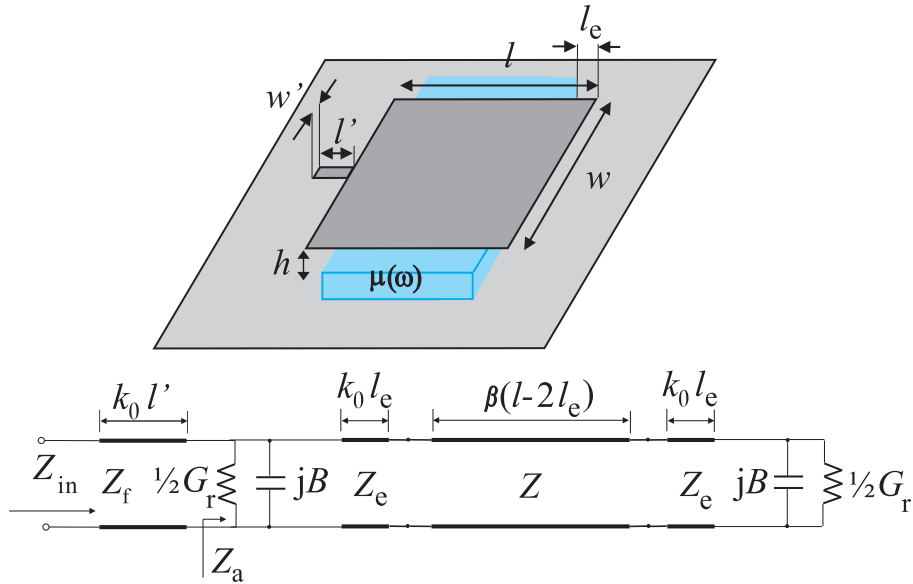, width=12cm} \caption{Schematic
illustration of the antenna geometry and the equivalent circuit of a
strip fed antenna.} \label{sche}
\end{figure}

In this section we will derive
expressions for the input impedance $Z_{\rm in}$ and unloaded
quality factor $Q_0$ of an arbitrary size strip fed patch antenna
(lying on top of a large, non-resonant ground plane). The impedance
level seen directly at the patch edge is usually very high causing
difficulties in impedance matching of the antenna. The strip feeding
allows one to conveniently tune the impedance level seen at the
feeding point, thus it offers a possibility to tune the matching in
different scenarios. A schematic illustration of the analyzed
antenna structure and the equivalent TL-model are presented in
Fig.~\ref{sche}. The characteristic impedance of a wide microstrip
line reads \cite{Wheeler_adm}: \e Z =
\frac{\eta_0h}{w}\sqrt{\frac{\mu_{\rm eff}}{\E_{\rm eff}}}, \l{Z} \f
where $\eta_0$ is the wave impedance in free space, $h$ is the
height of the substrate (assumed to be the same as the height of the
patch from the ground plane), $l$ is the length of the patch, and
$\mu_{\rm eff}$ and $\E_{\rm eff}$ are the effective substrate
material parameters.
In the present model we neglect the effect of the material filling
to the radiation conductance and the shunt susceptance representing
the open end extension of the TL. This is possible if the material
filling leaves the ends of the patch free so that there is a short
(about the patch height) section of the empty TL just near the
radiating edges (in the present analysis the length of the empty
section $l_{\rm e} = h/2$). With this assumption the radiation
conductance and shunt susceptance read \cite{Bahl,Balanis}: $$G_{\rm
r} = \frac{1}{90}\bigg{(}\frac{w}{\lambda_0}\bigg{)}^2, \quad w <
0.35\lambda_0. $$
$$
G_{\rm r} = \frac{1}{120}\frac{w}{\lambda_0} - \frac{1}{60\pi^2},
\quad 0.35\lambda_0 \leq w \leq 2\lambda_0,$$ \e G_{\rm r} =
\frac{1}{120}\frac{w}{\lambda_0}, \quad w>2\lambda_0. \label{Gr} \f
\e B = 2\pi\frac{\Delta{l}}{Z\lambda_0}. \label{B} \f Above $w$ is
the width of the patch, and $\lambda_0$ is the wavelength in free
space. In eq.~(\ref{B}) $\Delta{l}$ is the open end extension of the
TL (with a wide microstrip $\Delta{l}\approx{h/2}$). The propagation
factor in the antenna part of the TL is of the following form \e
\beta = k_0\sqrt{\mu_{\rm eff}\E_{\rm eff}}, \label{beta} \f where
$k_0$ is the free space wave number. For the input impedance of the
antenna we can write \e Z_{\rm in} = Z_{\rm f}\frac{Z_{\rm a} +
jZ_{\rm f}\tan{k_0l'}}{Z_{\rm f} + jZ_{\rm a}\tan{k_0l'}},
\label{Zin} \f where $Z_{\rm a}$ is the impedance seen from the
reference plane of the first radiating slot and $l'$ is the length
of the feed strip. The impedance of the (narrow) feed line is
\cite{Wheeler_narrow} \e Z_{\rm f} =
\frac{42.5}{\sqrt{2}}\ln\bigg{[}1 +
\frac{4h}{w'}\bigg{(}\frac{8h}{w'} +
\sqrt{\bigg{(}\frac{8h}{w'}\bigg{)}^2 + \pi^2}\bigg{)} \bigg{]},
\label{Zfl} \f where $w'$ is the width of the feed strip. The
impedance level of the feeding probe (connected to the end of the
strip) is assumed to be $Z_0=50$, and the reflection coefficient is
defined as \e \rho = \frac{Z_{\rm in} - Z_0}{Z_{\rm in} + Z_0}.
\label{roo} \f To relate the unloaded quality factor to the return
loss characteristics, we regard the antenna as a parallel-resonant
RLC circuit in the vicinity of the fundamental resonant frequency
\cite{Pues}. The coupling coefficient $T$ describes the quality of
the matching between the source and the resonator and is defined as
\e T = \frac{Y_{\rm f}}{G}, \label{T} \f where $Y_{\rm f}$ is the
admittance of the feed line, and $G=1/R$ is the conductance of the
parallel resonator. Let us assume that the criterion for the maximum
VSWR inside of the band of interest is \e {\rm VSWR} \leq S.
\label{VSWR} \f In this case the fractional bandwidth can be
expressed as \cite{Pues} \e BW = \frac{1}{Q_0}\sqrt{\frac{(TS - 1)(S
- T)}{S}}, \label{bwr} \f where $Q_0$ is the unloaded quality factor
of the parallel resonator. Knowing the dispersive behavior of the
input return loss, $Q_0$ can be solved from (\ref{bwr}).

A half-wavelength patch antenna fed directly at the patch edge
operates as a parallel resonant circuit when $l=\lambda/2$. The feed
strip can be interpreted as an additional inductance placed in
series with the rest of the circuit. This inductance will create a
series resonance above the parallel resonance, and usually the
antenna operates in between the series and the parallel resonance.

The dispersive behavior of the metasolenoid array is of the
following form \cite{Maslovski} (see Fig.~\ref{meta_meas}): \e
\mu_{\rm eff} = 1 - V_{\rm r}\frac{j\omega\mu_0\Lambda}{Z_{\rm
tot}d}, \label{permeability} \f where $\Lambda=ab$ is the
cross-section area of the metasolenoid, and $V_{\rm r}$ is a
coefficient taking into account realistic bulk concentration. The
total impedance of the metasolenoid reads \cite{Maslovski}: \e
Z_{\rm tot} = j\omega{L_{\rm eff}} + \frac{1}{j\omega{C_{\rm eff}}}
+ R_{\rm eff}. \label{Ztot} \f The effective permittivity of the
metasolenoid array does not equal to that of the host substrate
relative permittivity ($\E_{\rm eff}\neq\E_{\rm r}^{\rm s}$). The
electric resonant behavior of the metasolenoid is very weak
(compared to the magnetic resonant behavior) \cite{Maslovski}, thus
a good approximation for the permittivity is a frequency independent
expression \e \E_{\rm eff} = m(V_{\rm r})\E_{\rm eff}^{\rm s},
\label{eps_eff} \f where $m(V_{\rm r})$ is a multiplication factor
depending on the bulk concentration ($m(V_{\rm r})>1$), and $\E_{\rm
eff}^{\rm s}$ is the effective permittivity of a host substrate
having relative permittivity $\E_{\rm r}$.

\section{Impedance bandwidth properties}

\subsection{Dispersion-free material parameters}

In the first case we load the volume between the patch element and
the ground plane with hypothetical, lossless substrates having
static material parameters. The following loading scenarios are
considered (values are for the substrate material parameters): 1)
$\E_{\rm eff}=\mu_{\rm eff}=1$, 2) $\E_{\rm eff}=1$, $\mu_{\rm
eff}=8$, 3) $\E_{\rm eff}=6.75$, $\mu_{\rm eff}=1$, 4) $\E_{\rm
eff}=\mu_{\rm eff}=2.65$. The empty antenna has dimensions (see
Fig.~\ref{sche}) $l=w=48.5$ mm, $h=4$ mm, $l'=15$ mm, $w'=8.4$ mm.
With the aforementioned dimensions the empty antenna resonates at
3.0 GHz. The dimensions of the patch loaded with high-$\mu$,
high-$\E$, and magneto-dielectric substrates are the following:
$l=w=22.5$ mm, $h=4$ mm, $l'=15$ mm, $w'=1.5$ mm (high-$\mu$
substrate), $w'=1.2$ mm (high-$\E$ substrate and magneto-dielectric
substrate). Feed strip width $w'$ has been left as a free parameter
and is used to tune the quality of coupling in different loading
scenarios. With all the loading scenarios $T$ has been tuned to
$T=T_{\rm opt}=1/2(S + 1/S)$ \cite{Pues} by varying $w'$. This
allows us to neglect the square root term in (\ref{bwr}) when
comparing unloaded quality factors (note that coupling fine tuning
is a common procedure in practical antenna design where the goal
usually is to maximize the impedance bandwidth).

\begin{figure}[t!]
\centering \epsfig{file=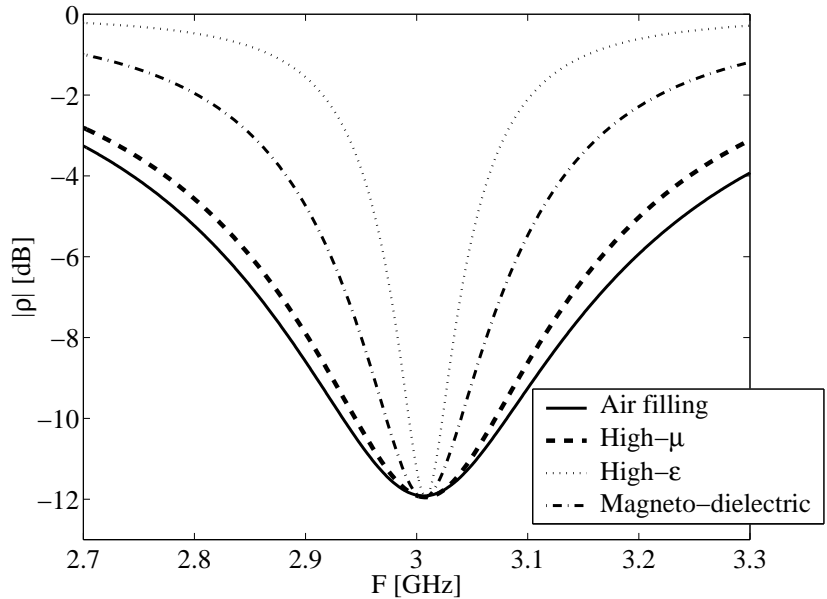, width=11cm}
\caption{Calculated reflection coefficient with different material
fillings. Dispersion-free $\mu_{\rm eff}$.} \label{S11_Hansen}
\end{figure}

Fig.~\ref{S11_Hansen} shows the calculated reflection coefficient
for a patch filled with different substrates. The main calculated
parameters are gathered in Table \ref{Hansen_t}. The unloaded
quality factor obtained from eq.~(\ref{bwr}) has been calculated
using --6 dB matching criterion. We can observe from
Fig.~\ref{S11_Hansen} and Table~\ref{Hansen_t} that the impedance
bandwidth properties behave similarly as predicted by Hansen and
Burke \cite{Hansen}: Dispersion-free high-$\mu$ materials allow size
miniaturization while practically retaining the impedance bandwidth
whereas when using high-$\E$ materials the impedance bandwidth
suffers significantly. We can also observe from
Fig.~\ref{S11_Hansen} that increase in permittivity rather quickly
increases the unloaded quality factor and narrows the impedance
bandwidth (case ``magneto-dielectric'').

\begin{table}[b!]
\centering \caption{The calculated impedance bandwidth properties.
Dispersion-free $\mu_{\rm eff}$.} \label{Hansen_t}
\begin{tabular}{|l|c|c|c|}
\hline Loading & $V$  & $BW\bigg{|}_{\rm -6 dB}$ & $Q_0$ \\
 & cm$^3$ & percent & \\
\hline
Air filling& 9.4 & 12.3 & 10.9 \\
High-$\mu$ & 2.0 & 10.5 & 12.8 \\
High-$\E$ & 2.0 & 2.5 & 53.2 \\
Magneto-dielectric & 2.0 & 5.6 & 24.0 \\
\hline
\end{tabular}
\smallskip
\end{table}

\subsection{Dispersive material parameters}

\begin{figure}[t!]
\centering \epsfig{file=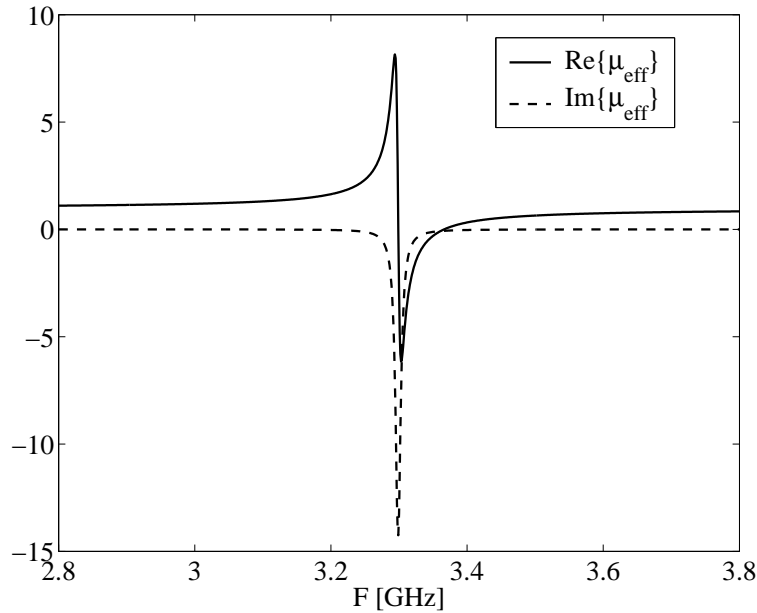, width=10.0cm} \caption{Dispersive
behavior of $\mu_{\rm eff}$ of a practically realizable substrate.}
\label{mu}
\end{figure}

In most of the works found in the literature for antenna
miniaturization using artificial magneto-dielectric substrates
(e.g.~\cite{Mosallaei_Pisa, Yoon}) the frequency dispersion of the
substrate has been neglected, and scalar constant-permeability
assumption is used. In other words, it is assumed that the behavior
of the loaded antenna is the same if instead of embedding the full
frequency dispersion to the analysis, one picks up the values for
$\mu_{\rm eff}$ at the operational frequency of the loaded antenna.
The purpose of this subsection is to check the validity of scalar
constant-permeability assumption when the substrate obeys Lorentzian
type dispersion for $\mu_{\rm eff}$.

We start the analysis with calculating the impedance bandwidth
properties when the antenna is loaded with a substrate whose
material parameters correspond to practically realizable values. The
dispersive behavior of $\mu_{\rm eff}$ of a practically realizable
magneto-dielectric substrate is shown in Fig.~\ref{mu}. We can see
that Re$\{\mu_{\rm eff}\}=1.21$ at 3.0 GHz. We estimate the
effective permittivity of the substrate to be $\E_{\rm eff}=8.5(1 -
j0.001)$ \cite{Sim}. For the reference substrate we tune the value
of the relative permittivity so that the resonant frequency of the
dielectrically loaded antenna coincides with the resonant frequency
of the magneto-dielectrically loaded antenna. The corresponding
value for the relative permittivity of the reference dielectric
substrate is found to be $\E_{\rm r}^{\rm ref} = 10.1(1 - j.001)$.
The dimensions of the loaded patches are the following: $l=w=19.3$
mm, $h=4$ mm, $l'=15$ mm, $w'=0.9$ mm (magneto-dielectric
substrate), $w'=0.75$ mm (reference dielectric substrate). To better
see the possible effect of frequency dispersion, we will also
consider a loading scenario in which the dispersive
magneto-dielectric substrate is replaced with a substrate having
\emph{dispersion-free} material parameters $\mu_{\rm eff}=1.21(1 -
j0.0024)$ (picked up from the dispersion curve at the operational
frequency of the loaded antenna), $\E_{\rm eff}=8.5(1 - j0.001)$.

\begin{figure}[t!]
\centering \epsfig{file=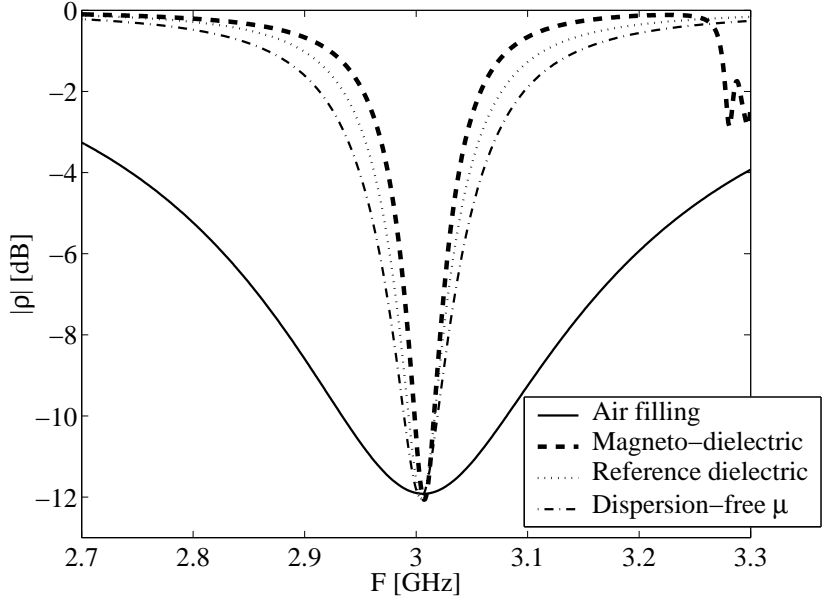, width=11cm}
\caption{Calculated reflection coefficient with different material
fillings. Practically realizable example.} \label{S11_real}
\end{figure}

Fig.~\ref{S11_real} shows the calculated reflection coefficient with
different material fillings. The main calculated parameters are
gathered in Table \ref{Real_t}. The obtained result indicates that
practically realizable magneto-dielectric substrate offers no
advantages over high-permittivity dielectrics. We notice that in
this particular case the frequency dispersion of the substrate is
rather weak. However, this weak dispersion added to significantly
high $\E_{\rm eff}$ (as compared to $\mu_{\rm eff}$) leads to poor
impedance bandwidth properties. If the realistic dispersive behavior
of the magneto-dielectric substrate is replaced with a scalar
constant-permeability, the TL-model predicts wider impedance
bandwidth with magneto-dielectrics than with pure high-permittivity
dielectrics, also in the case when Re$\{\mu_{\rm eff}\}\ll$
Re$\{\E_{\rm eff}\}$. This result is in line with the general
opinion in the literature (based on static material parameters).
However, when thinking of practical antenna design, dispersion-free
assumption gives a wrong impression on the applicability of the
magneto-dielectric substrate.

\begin{table}[b!]
\centering \caption{The calculated impedance bandwidth
characteristics. Practically realizable example.} \label{Real_t}
\begin{tabular}{|l|c|c|c|}
\hline Loading & $V$  & $BW\bigg{|}_{\rm -6 dB}$ & $Q_0$ \\
 & cm$^3$ & percent & \\
\hline
Air filling& 9.4 & 12.3 & 10.9 \\
Magneto-dielectric & 1.5 & 1.4 & 93.7 \\
Reference dielectric & 1.5 & 1.9 & 69.7 \\
Dispersion-free $\mu$ & 1.5 & 2.5 & 54.2 \\
\hline
\end{tabular}
\smallskip
\end{table}

Next, we consider the role of high Re$\{\E_{\rm
eff}\}$/Re$\{\mu_{\rm eff}\}$ added to dispersive $\mu_{\rm eff}$.
To reveal whether or not practically realizable, dispersive
magneto-dielectric substrate \emph{with a low value for}
Re$\{\E_{\rm eff}\}$ offers any advantage over (dispersion-free)
low-permittivity dielectrics, we will load the antenna with
dispersive $\mu_{\rm eff}$ (corresponding to a practically
realizable substrate) and assume that Re$\{\E_{\rm eff}\}$ of the
substrate is unity\footnote{In principe condition Re$\{\E_{\rm
eff}\}=1$ can be realized e.g.~using the wire medium
\cite{Maslovski_wires, Belov}. However, condition Re$\{\E_{\rm
eff}\}$ is fulfilled only at one frequency, and the realized
effective permittivity is necessarily dispersive.}. The results for
this loading scenario are again challenged against the results
obtained with pure dielectrics offering the same size reduction. The
size of the loaded antenna is $l=w=44.2$ mm, the feed strip width in
both loading scenarios is $w'=2.0$ mm. The reference dielectric
substrate has $\E_{\rm r}^{\rm ref} = 1.19(1 - j.001)$.

\begin{figure}[t!]
\centering \epsfig{file=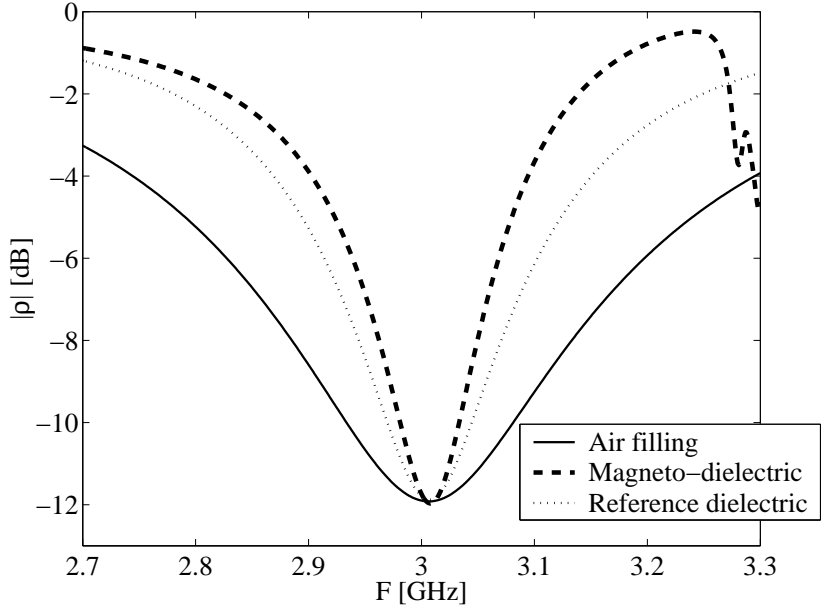, width=11cm}
\caption{Calculated reflection coefficient with different material
fillings. $\E_{\rm eff}=1$ for the magneto-dielectric substrate.}
\label{S11_weak_mu}
\end{figure}

Fig.~\ref{S11_weak_mu} shows the calculated reflection coefficient
with different material fillings. The main calculated parameters are
gathered in Table \ref{weak_mu}. The calculated result indicates
that there is no advantage in using magneto-dielectric substrates
which obey the Lorentzian type dispersion rule, even if Re$\{\E_{\rm
eff}\}=1$, and the dispersion in $\mu_{\rm eff}$ is weak. Thus, a
high value for Re$\{\E_{\rm eff}\}$ compared to Re$\{\mu_{\rm
eff}\}$ is not the factor which eventually deteriorates the
performance of the magneto-dielectric substrate.

The above discussion holds for artificial magneto-dielectric
(magnetic) substrates whose static value for Re$\{\mu_{\rm eff}\}=1$
(Lorentzian type dispersion rule). According to the Rozanov limit
\cite{Rozanov} for the thickness to bandwidth ratio of radar
absorbers, the thickness of the absorber at microwave frequencies
(with a given reflectivity level) is bounded by the \emph{static}
value of magnetic permeability of the absorber. It is interesting to
apply Rozanov's observation also in antenna miniaturization, and see
if a weakly dispersive magneto-dielectric substrate with static
value Re$\{\mu_{\rm eff}\}>1$ offers noticeable improvement in the
impedance bandwidth properties. For example, hexaferrite, or
composite materials containing ferromagnetic inclusions can be used
to produce the condition Re$\{\mu_{\rm eff}\}>1$ at zero frequency.
Even though the dispersive behavior of $\mu_{\rm eff}$ of ferrites
(or composites containing ferromagnetic inclusions) is usually
described by the Debye type dispersion rule, we continue to study
the Lorentzian type dispersion rule for the sake of clarity. Thus,
we assume that it is possible to build a composite material obeying
the Lorentzian type dispersion with static Re$\{\mu_{\rm eff}\}=2$.
Otherwise the dispersive behavior of $\mu_{\rm eff}$ is the same as
depicted in Fig.~\ref{mu}. Moreover, $\E_{\rm eff}=1$.

\begin{table}[b!]
\centering \caption{The calculated impedance bandwidth
characteristic. $\E_{\rm eff}=1$ for the magneto-dielectric
substrate} \label{weak_mu}
\begin{tabular}{|l|c|c|c|}
\hline Loading & $V$  & $BW\bigg{|}_{\rm -6 dB}$ & $Q_0$ \\
 & cm$^3$ & percent & \\
\hline
Air filling& 9.4 & 12.3 & 10.9 \\
Magneto-dielectric & 7.8 & 4.3 & 18.5 \\
Reference dielectric & 7.8 & 6.3 & 12.6 \\
\hline
\end{tabular}
\smallskip
\end{table}

\begin{figure}[t!]
\centering \epsfig{file=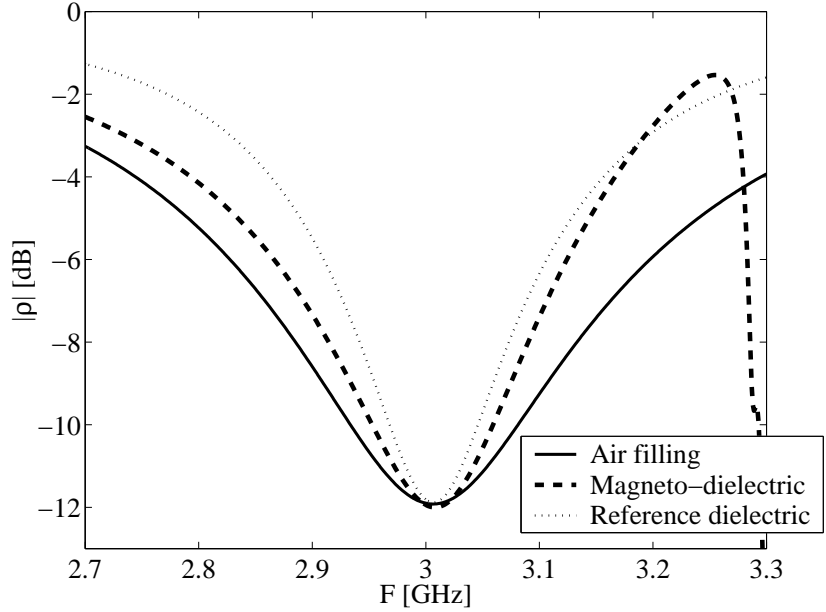, width=11cm} \caption{Calculated
reflection coefficient with different material fillings. Static
value for Re$\{\mu_{\rm eff}\}=2$, $\E_{\rm eff}=1$ for the
magneto-dielectric substrate.} \label{S11_mod}
\end{figure}

The dimensions of the loaded patches are the following: $l=w=33.5$
mm, $h=4$ mm, $l'=15$ mm, $w'=4.4$ mm (for both substrates). The
reference dielectric substrate has $\E_{\rm r}^{\rm ref} = 2.14(1 -
j.001)$. Fig.~\ref{S11_mod} shows the calculated reflection
coefficient with different material fillings. The main calculated
parameters are gathered in Table \ref{mod}. We can observe, that the
frequency dispersion in $\mu_{\rm eff}$ is outweighted by the
increased static value for Re$\{\mu_{\rm eff}\}$. Thus it seems that
weak frequency dispersion modulating a static value Re$\{\mu_{\rm
eff}\}>1$ leads to similar desired performance as hypothetical,
high-permeability dispersion-free substrates (of course the effect
of weak frequency dispersion is seen in the slightly deteriorated
impedance bandwidth as compared to dispersion-free assumption).

\begin{table}[b!]
\centering \caption{The calculated impedance bandwidth
characteristic. Static value for Re$\{\mu_{\rm eff}\}=2$, $\E_{\rm
eff}=1$ for the magneto-dielectric substrate} \label{mod}
\begin{tabular}{|l|c|c|c|}
\hline Loading & $V$  & $BW\bigg{|}_{\rm -6 dB}$ & $Q_0$ \\
 & cm$^3$ & percent & \\
\hline
Air filling& 9.4 & 12.3 & 10.9 \\
Magneto-dielectric & 5.0 & 8.6 & 15.5 \\
Reference dielectric & 5.0 & 6.5 & 20.5 \\
\hline
\end{tabular}
\smallskip
\end{table}

\subsection{Relative radiation quality factor}

Let us next derive an expression explicitly explaining the negative
effect of frequency dispersion. We consider a $\lambda/2$ long
section of a transmission line (which models a resonant patch
element) filled with a certain material having material parameters
$\mu,\E$, in general dispersive, but lossless. A standing wave is
formed under the patch. To find the energy stored under the patch we
use the known relation for the volume density of electromagnetic
field energy: \e w_{\rm em} = {\E_0 \partial(\omega\E) \over
\partial\omega}{E_{\rm m}^2\over 4} + {\M_0 \partial(\omega\M)
  \over \partial\omega}{H_{\rm m}^2\over 4},
\f where $E_{\rm m}$ and $H_{\rm m}$ are the amplitudes (real) of
the electric and magnetic fields. This formula holds for low-loss
materials. The amplitudes change along the patch with respect to the
standing wave pattern. The total energy stored under the patch can
be found by integrating $w_{\rm em}$ over the volume under the
patch. For a resonant patch this results in \e W_{\lambda/2} = {\pi
Y \over
16\omega}\bigg{(}\frac{1}{\mu}\frac{\partial(\omega\mu)}{\partial\omega}
+ \frac{1}{\E}\frac{\partial({\omega\E})}{\partial{\omega}}\bigg{)}
U_{\rm max}^2. \f Here $U_{\rm max}$ is the voltage amplitude (real)
at the voltage maximum of the standing wave, and $Y$ is the
characteristic admittance of the patch (given by eq.~\r{Z}).
Practically, $U_{\rm max}$ is the voltage at the end of the patch.

\begin{figure}[b!]
\centering \vskip 5mm \epsfig{file=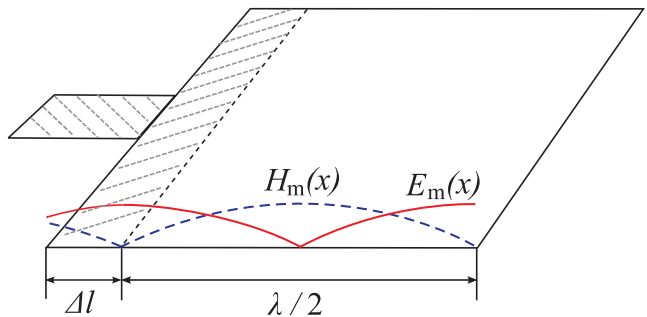,width=0.5\textwidth}
\caption{Patch operating slightly above its resonant frequency. The
  stroked areas represent the areas where additional reactive energy
  is stored.}
\label{energy}
\end{figure}

In Section \ref{tl_model_formulation} we mentioned that usually a
patch antenna is fed by a narrower microstrip line which introduces
additional series inductance so that the whole system becomes
double-resonant. The second, series resonance appears at a bit
higher frequency when the additional inductive impedance of the
narrow microstrip is compensated by the capacitive impedance of the
patch above its parallel resonance. This situation is outlined in
Fig.~\ref{energy}.
Now the total energy stored under the patch is the sum \e W_{\rm
patch} = W_{\lambda/2} + W_{\Delta l}, \f where $W_{\Delta l}$ is
the additional energy stored under the part of the patch that
exceeds $\lambda/2$ length. If $\Delta l \ll \lambda/4$ we can
neglect the magnetic energy stored in this segment. Then, we get \e
W_{\Delta l} \approx {\pi Y \over 2\omega\E} {\partial(\omega
\E)\over\partial\omega}\left({\Delta l\over
  \lambda}\right)U_{\rm max}^2.
\f

The amount of energy stored in the short narrow feed operating as an
effective inductor can be found as follows. First, we notice that
for an inductor the stored energy can be expressed in terms of the
current and voltage amplitudes on the inductor: \e W_L = {U^{L}_{\rm
m} I^{L}_{\rm m}\over 4\omega}. \f At the resonance the voltage
amplitudes on the reactive elements of the opposite nature are the
same (if the patch susceptance dominates over the radiation
conductance). Therefore, the voltage amplitude on the inductor
equals the voltage amplitude at the input of the patch, and since
the elements are connected in series the current through the
inductor is the input current of the patch: \e U^L_{\rm m} \approx
U_{\rm max}, \quad I^L_{\rm m} \approx 2\pi Y \left(\Delta
l\over\lambda\right) U_{\rm max}. \f From here \e W_L = {\pi Y\over
2\omega}\left(\Delta l\over\lambda\right) U_{\rm max}^2. \f Finally,
the total energy stored in the whole system at the series resonance
is \e W_{\rm tot} = W_{\lambda/2} + W_{\Delta l} + W_L \ge
W_{\lambda/2}. \l{total_energy} \f

From above the quality factor of a resonant $\lambda/2$ patch (which
 is the lowest possible quality factor in view of \r{total_energy})  can be
 found as
 \e Q_{\rm r} = {\omega W_{\lambda/2}\over P_{\rm r}}=\frac{\pi{Y}}{8G_{\rm
r}}\bigg{(}\frac{1}{\mu}\frac{\partial(\omega\mu)}{\partial\omega} +
\frac{1}{\E}\frac{\partial({\omega\E})}{\partial{\omega}}\bigg{)}.
\label{Qdisp} \f
Here $G_{\rm r}$ is the radiation conductance and $P_{\rm r} = G_{\rm
  r} U_{\rm
  max}^2/2$ is the radiated power (radiation happens at the ends of
  the patch). For a patch antenna having the same dimensions and
loaded with a reference, dispersion-free magneto-dielectric material
we have \e Q_{\rm r}^{\rm ref} = \frac{\pi{Y^{\rm ref}}}{4G_{\rm
r}}, \label{refQ} \f where $Y^{\rm ref}$ is the characteristic
admittance of the reference antenna. For now on we only consider
resonant $\lambda/2$ patches. The following holds for the ratio
between the radiation quality factors: \e \frac{Q_{\rm r}}{Q_{\rm
r}^{\rm ref}} = \frac{1}{2\mu}\sqrt{\frac{\E\mu_{\rm
ref}}{\mu\E_{\rm
ref}}}\bigg{(}\frac{1}{\mu}\frac{\partial(\omega\mu)}{\partial\omega}
+ \frac{1}{\E}\frac{\partial({\omega\E})}{\partial{\omega}}\bigg{)}.
\label{ratio} \f Since the two antennas resonate at the same
frequency we have \e \mu\E = \mu_{\rm ref}\E_{\rm ref}. \label{rel}
\f Further, if we consider the reference material to be purely
dielectric ($\mu_{\rm ref}=1$) eq.~(\ref{ratio}) simplifies to \e
\frac{Q_{\rm r}}{Q_{\rm r}^{\rm ref}} =
\frac{1}{2\mu}\bigg{(}\frac{1}{\mu}\frac{\partial(\omega\mu)}{\partial\omega}
+ \frac{1}{\E}\frac{\partial({\omega\E})}{\partial{\omega}}\bigg{)}.
\label{ratio_2} \f To be consistent with the analysis presented in
the paper, we will further assume that $\E\approx$ const. In this
case we obtain \e \frac{Q_{\rm r}}{Q_{\rm r}^{\rm ref}} =
\frac{1}{2\mu}\bigg{(}\frac{1}{\mu}\frac{\partial(\omega\mu)}{\partial\omega}
+ 1\bigg{)}. \label{ratio_3} \f Furthermore, we assume the
dispersion of $\mu$ to be covered by the general (lossless)
Lorentzian type dispersion rule \e \mu = 1 +
\frac{A\omega^2}{\omega_0^2 - \omega^2}, \label{lor} \f where $A$ is
the amplitude factor ($0<A<1$) and $\omega_0$ is the undamped
frequency of the zeroth pole pair.

\begin{figure}[t!]
\centering \epsfig{file=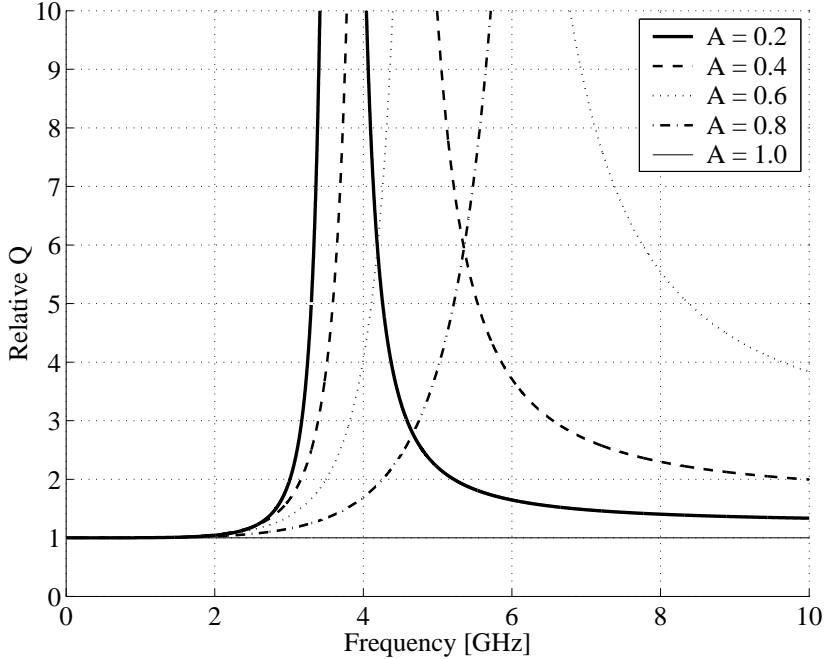, width=11cm} \caption{The relative
quality factor. Static Re$\{\mu\}=1$.} \label{rel_Q}
\end{figure}
\begin{figure}[t!]
\centering \epsfig{file=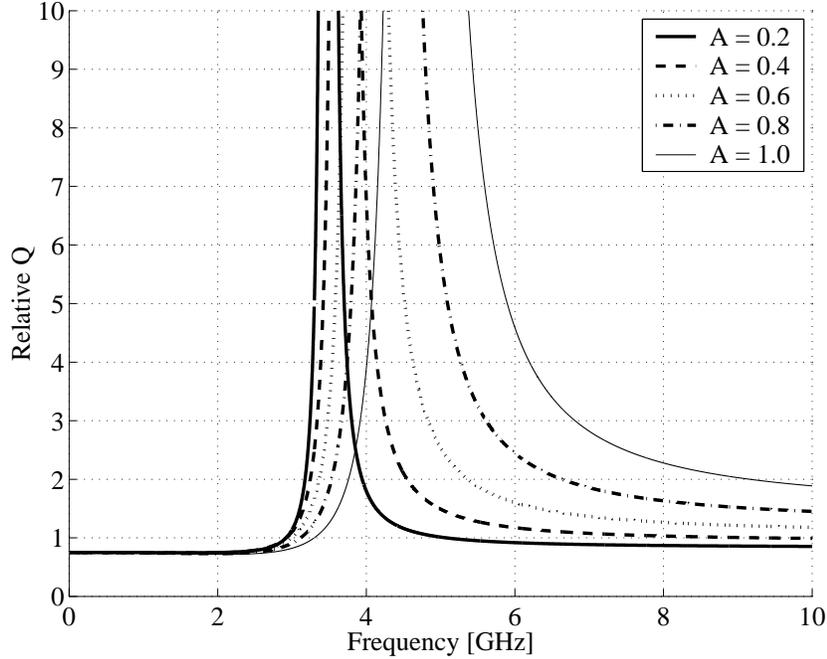, width=11cm} \caption{The
relative quality factor. Static Re$\{\mu\}=2$.} \label{rel_Q2}
\end{figure}

The relative quality factor (eq.~(\ref{ratio_3})) is presented with
different amplitude coefficients in Fig.~\ref{rel_Q} ($f_0$=3.3
GHz). We can observe that if the static value for Re$\{\mu\}=1$,
substrate with Lorentzian type dispersion of $\mu$ leads always to
larger $Q_{\rm r}$ than pure dielectrics offering the same size
reduction (except in the limiting case with $A=1$). This result is
in line with the results presented earlier calculated using the
TL-model. However, if the static Re$\{\mu\}=2$, there exist
frequency bands, over which the magneto-dielectric substrate
outperforms pure dielectrics in what comes to minimized $Q_{\rm r}$,
Fig.~\ref{rel_Q2}. Only close to $\omega_0$ the strong frequency
dispersion outweights the effect of increased static Re$\{\mu_{\rm
eff}\}$. Again this result agrees with the results calculated using
the TL-model.

\begin{figure}[b!]
\centering \epsfig{file=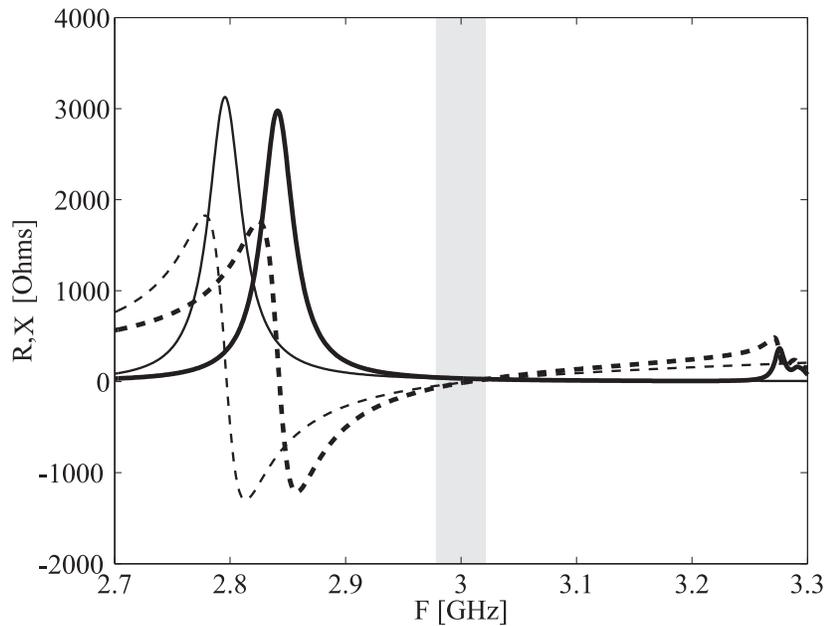, width=11cm} \caption{Calculated
input impedance for the practically realizable case. Thick lines are
for the magneto-dielectric substrate, thin lines for
high-permittivity substrate. The grey region denotes the --6 dB
matching band of the antennas.} \label{Zin}
\end{figure}

Physical understanding of the effect of frequency dispersion in the
substrate material parameters can be achieved also by inspecting the
input impedance of antennas loaded by dispersive magneto-dielectrics
and pure dielectrics leading to the same size reduction.
Fig.~\ref{Zin} presents the calculated input impedance when the
antenna has been loaded with a practically realizable
magneto-dielectric substrate and the corresponding reference
substrate (the input return loss for this case is presented in
Fig.~\ref{S11_real}). The known definition for the quality factor of
a resonator near its parallel resonance reads: \e Q =
\frac{\omega{W}}{P} =
\frac{\omega}{2G}\frac{\partial{B}}{\partial\omega} \label{Q}. \f
Above, $\omega$ is the angular frequency, $W$ denotes the amount of
stored energy in the volume defined by the near fields of the
antenna, $P$ is the power dissipated during one cycle, $B$ is the
susceptance of the resonator, and $G$ represents loss conductance.
In the case of a series resonance, $B$ is replaced by the reactance
$X$, and $G$ by the loss resistance $R$. Physically the behavior of
the magneto-dielectric substrate is understandable from
Fig.~\ref{Zin} and eq.~(\ref{Q}): We can see that the input
reactance
changes more strongly at the resonance in the case of
magneto-dielectric loading. This is due to the material resonance
appearing at 3.3 GHz. In terms of electromagnetic energy, this rapid
change in the input reactance corresponds to increased effect of
reactive near fields leading to increased amount of stored energy,
thus to increased quality factor.

\section{Experimental demonstration}

\begin{figure}[b!]
\centering \epsfig{file=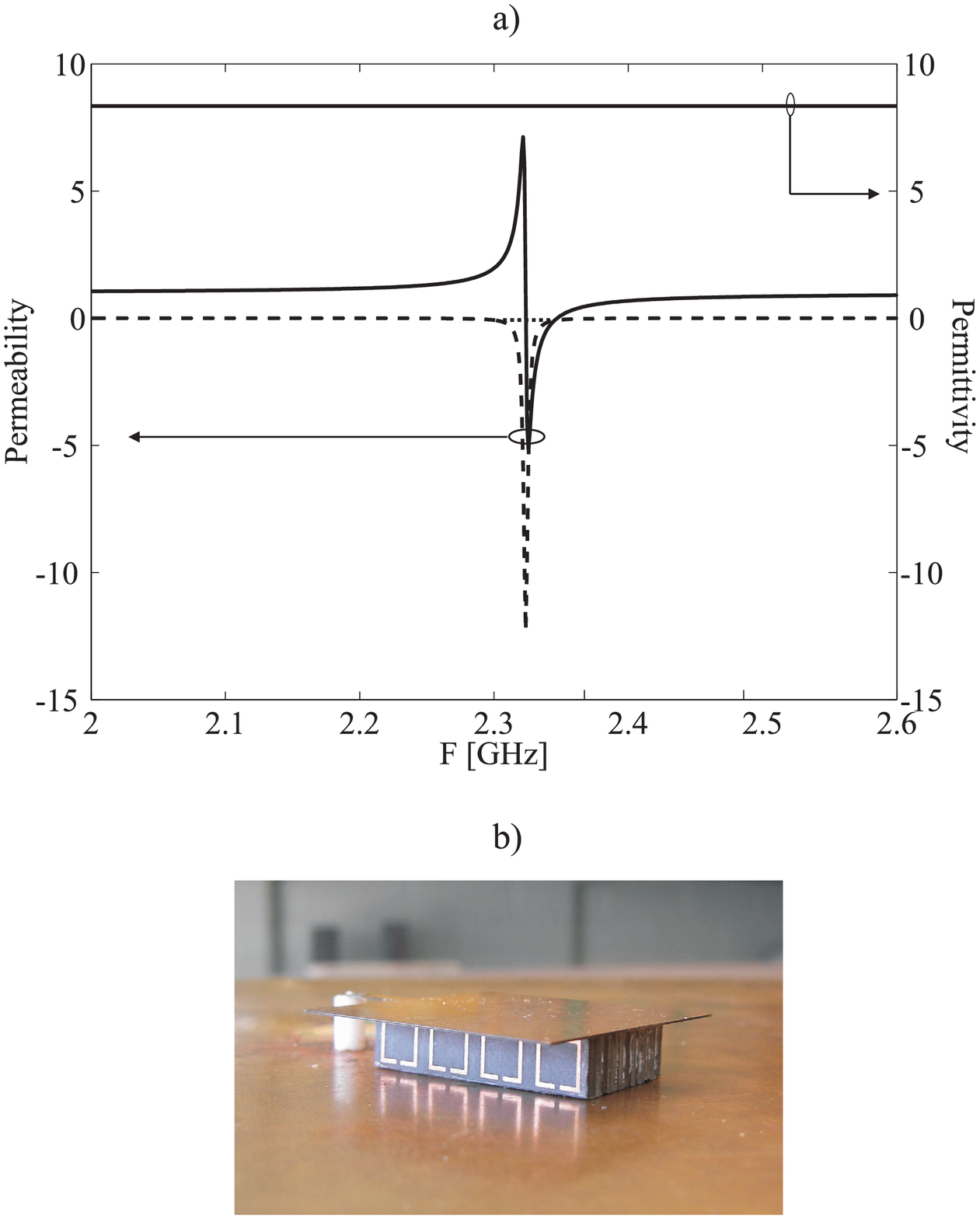, width=11cm} \caption{a) The
estimated material parameters for the implemented magneto-dielectric
substrate. Solid lines for the real parts, dashed lines for the
imaginary parts. b) The manufactured prototype antenna.}
\label{proto}
\end{figure}

We will manufacture a prototype antenna and load the volume under
the antenna with an array of metasolenoids. For reference, the
impedance bandwidth will also be measured when the volume under the
antenna is loaded with high-permittivity dielectrics leading to the
same size reduction. The estimated dispersive behavior of the the
manufactured metasolenoid array and a photograph of the prototype
antenna are shown in Fig.~\ref{proto}a and \ref{proto}b,
respectively. At the operational frequency of the loaded antenna
($F$ = 2.07 GHz) we estimate that Re$\{\mu_{\rm eff}\}=1.25$ and
Re$\{\E_{\rm eff}\}=8.5$ (the host substrate for the metasolenoids
is Rogers R/T Duroid 5870). The dimensions of the empty antenna are
$l=w=66$ mm, $h=7.5$ mm, $l'=5$ mm, $w'=4$ mm. The dimensions of the
loaded antenna are $l=w=35$ mm, $h=7.5$ mm, $l'=10$ mm, $w'=3$ mm.
The relative permittivity for the the reference dielectric leading
to the same size reduction is $\E_{\rm r}^{\rm ref} = 10.8(1 -
j.0037)$.

\begin{figure}[b!]
\centering \epsfig{file=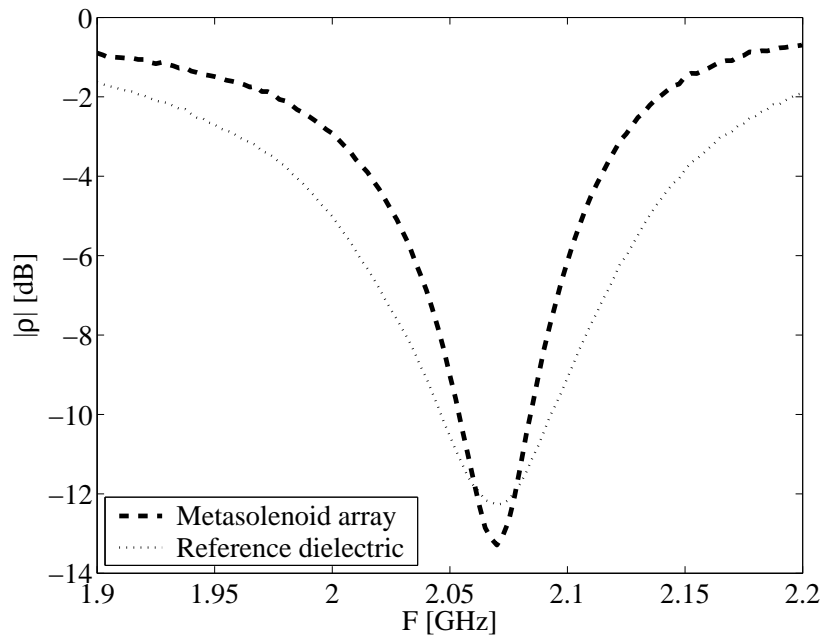, width=11cm} \caption{The
measured reflection coefficient with different material fillings.}
\label{meas}
\end{figure}

Fig.~\ref{meas} and Table \ref{meas_table} show the measured
reflection coefficient and gather the main measured parameters. The
radiation efficiency $\eta_{\rm rad}$ has been measured using the
Wheeler cap method. The unloaded quality factor obtained from
eq.~(\ref{bwr}) has been calculated using --6 dB matching criterion.
We can see that the impedance bandwidth behaves according to the
analysis presented in the paper: Practically realizable
magneto-dielectric substrate does not improve the impedance
bandwidth in antenna miniaturization compared to pure dielectrics.

\section{Conclusion}

In the present paper we have systematically studied the effect of
artificial magneto-dielectric substrates on the impedance bandwidth
properties of microstrip antennas. Using the transmission-line model
we have reproduced the known results for antenna miniaturization
using static material parameters. Frequency dispersion of the
substrate has been embedded into the analysis, and it has been shown
that with artificial magneto-dielectric substrates obeying the
Lorentzian type dispersion for $\mu_{\rm eff}$, frequency dispersion
can not be neglected in the analysis. A relation has been derived
for the ratio between radiation quality factors of ideally shaped
antennas loaded with dispersive magneto-dielectrics, and
dispersion-free reference dielectrics. The result shows that
dispersive magneto-dielectrics lead always to larger radiation
quality factor (Lorentzian type dispersive behavior) if static
Re$\{\mu_{\rm eff}\}=1$. The main observation on the negative effect
of frequency dispersion on the impedance bandwidth properties has
been experimentally validated.

\begin{table}[t!]
\centering \caption{The main measured parameters.}
\label{meas_table}
\begin{tabular}{|l|c|c|c|c|}
\hline Loading & $V$  & $BW\bigg{|}_{\rm -6 dB}$ & $Q_0$ & $\eta_{\rm rad}$ \\
 & cm$^3$ & percent &  & percent \\
\hline
Metasol.~array & 9.2 & 3.2 & 41.5 & 89 \\
Reference dielectric & 9.2 & 5.5 & 24.3 & 92 \\
\hline
\end{tabular}
\smallskip
\end{table}

\section*{Acknowledgement} Inspiring discussions with Prof.~Pertti
Vainikainen, Dr.~Jani Ollikainen, and Dr.~Murat Ermutlu are warmly
acknowledged.

\end{document}